\documentclass[aps,pra,preprint,showpacs,preprintnumbers,amsmath,amssymb,english,ednfloats]{revtex4-1}  
\usepackage[english]{babel}

\usepackage{here} 
\usepackage{subfigure}
\usepackage{graphicx}
\usepackage{dcolumn}
\usepackage{bm}
\usepackage{epstopdf}
\usepackage{epsfig}

\begin{document}
\title{A perturbative approach to self-phase modulation and self-steepening of short laser pulses propagating
in nonlinear media}
\author{F. Vidal} 
\affiliation{Institut national de la recherche scientifique-Centre \'energie,  mat\'eriaux et t\'el\'ecommunications, 1650 boul. Lionel Boulet, Varennes,  Qu\'ebec, Canada J3X 1S2}

\date{\today}

\begin{abstract}
The solution of the wave equation in the envelope approximation with temporal corrections for a laser pulse propagating in a medium where the Kerr effect, field ionization, and associated absorption take place, is obtained through a first-order perturbative approach. The closed-form expressions so obtained clarify the influence of the various terms of the equation on the laser amplitude and on the frequency generation as a function of the retarded time. Furthermore, they allow extracting scaling parameters which size the nonlinear effects. The results are illustrated quantitatively on the case of a femtosecond pulse focused in air with typical parameters.
\end{abstract}


\maketitle
\section{Introduction}
\label{Intro}
The propagation of intense short laser pulses in optically transparent media involves several linear and nonlinear effects which modulate the pulse’s amplitude and phase. Identifying these effects and their influence on the laser pulse characteristics is of great inteterst due to the real or potential use of such laser pulses in various fields of science and technology such as communications \cite{Knox2000}, medicine \cite{Juhasz1999}, material science \cite{Gatlass2008}, sensors \cite{Mihailov2017}, high-harmonic generation \cite{Krause1992}, X-ray production \cite{Dopp2018}, etc. In the past decades, this topic has also brought much attention from a theoretical point of view \cite{Couairon2007,Berge2007,Couairon2011,Kolesik2014}. While the numerical approach is the only universal means to deal with fully developed nonlinear effects, it provides only case-by-case solutions and it requires numerical schemes whose accuracy must be carefully controlled. It is often possible, however, to derive closed-form solutions of nonlinear problems in the perturbative regime, that is when the nonlinear effects are not fully developed. Such analytical solutions, besides providing quick answers and reference points for numerical calculations, though in restricted conditions, allow a better understanding of the specific influence of the various terms and effects included in the basic nonlinear equations. In addition, they can provide dimensionless scaling parameters which help to evaluate the magnitude of the nonlinear effects. In this work, we present such closed-form expressions for the problem of interest in the perturbative regime and we illustrate them on the particular case of a femtosecond laser pulse focused in ambient air.

\section{Model equation}
\label{ModEq}
The basic equation for the field amplitude in the envelope approximation has
been comprehensively discussed by Berg\'{e} et al. \cite{Berge2007}. A simplified form of
this equation reads \cite{Ward2003}:
\begin{align}
& i\partial _{z}A+\frac{1}{2k_{0}}\frak{T}^{-1}\nabla _{\perp }^{2}A-
\frac{1}{2}k^{\prime \prime }\partial _{\tau }^{2}A+k_{0}n_{2}\frak{T}\left(
|A|^{2}A\right) -\frac{k_{0}}{2n^2_0\rho _{c}}\frak{T}^{-1}\left( \rho
A\right)  \nonumber \\
& +i\frac{\beta }{2}|A|^{2K-2}A+i\frac{\sigma}{2}\rho A=0
\label{eqgen}
\end{align}
where $z$ is the propagation axis coordinate, $\tau =t-z/v_{g}$ is the
retarded time with $v_g$ the group velocity, $\frak{T}=1+\frac{i}{\omega_c}\partial _{\tau }$, with
$\omega_c$ the central frequency, and
$\rho(\tau) =\alpha _{K}\rho _{m}\int_{-\infty }^{\tau }|A|^{2K}d\tau ^{\prime }$
is the electron density generated by the laser pulse propagating in a medium where 
$\rho=0$ initially, $\rho _{m}$ is the particle density of the
propagation medium, here considered to be much larger than $\rho $. The field
amplitude $A(\mathbf{r},\tau)$ is normalized such that $|A|^{2}$ is
the intensity (irradiance) in Wm$^{-2}$. In order to allow analytical treatment, 
we approximate the ionization 
rate by a function of the form $W=\alpha _{K}|A|^{2K}$ around
some value of laser intensity $|\tilde{A}|^{2}$, where $\alpha _{K}$
and $K$ are fitted to $W(|A|^{2})$ around $|\tilde{A}|^{2}$.
For Keldysh parameters $\gamma_{K} > 1$, $K$ is the number of photons $N_{p}$ 
required to overcome the
ionization potential, otherwise $K$ is a real number such that $K<N_{p}$.
This approximation overestimates the ionization rate for $|A|^{2}$
higher or lower than $|\tilde{A}|^{2}$ since $K$ is a decreasing function 
of $|A|^2$ \cite{Schwarz2012}.

In Eq. (1), the meaning of the terms following the first one are: diffraction,
dispersion, Kerr effect, field ionization, absorption due to field ionization, 
and collisional (inverse bremsstrahlung) absorption by free electrons,
respectively. The constants
have their usual meaning: $k_{0}$ is the wave number in vacuum,
$n_{0}$ and $n_{2}$ are the 
linear and nonlinear indices of refraction, respectively,
$\rho_c$ is the plasma critical
density, $k^{\prime \prime }$ is the dispersion coefficient 
and $\sigma=e^2\nu_e/(m_e \varepsilon_0 \omega_c^2)$ in the limit 
$\nu_e/\omega_c \ll 1$ (with $e$ the electron charge, $m_e$ the electron rest mass, 
and $\varepsilon_0$ the vacuum permittivity)
is the collisional absorption cross section,
which depends on the electron collision frequency $\nu_e$. The operator 
$\nabla _{\perp }^{2}$ is the Laplacian in the transverse direction. 

To keep the problem tractable analytically, we make the following simplifications.
First, as dispersion introduces considerable complications in the perturbative
solution, it will not be considered. In fact, dispersion can be neglected in 
weakly dispersive media such as gases for sufficiently short propagation distances, 
as exemplified in Section V.
Second, we consider only the first order Taylor expansion
$\frak{T}^{-1}\approx 1-\frac{i}{\omega_c}\partial _{\tau }$.
Third, in order to decouple the transverse coordinates from
$z$ and $\tau$, we assume that the beam behaves as follows
close to the radial position $r=0$ \cite{Max1976}, 
\begin{equation}
A(z,r,\tau)=\frak{A}(z,\tau) \left[1-\left( \frac{1}{w^2(z)}-
i\frac{k_0}{2} \partial_z \ln(w(z)) \right)r^2+O(r^4)\right ]
\label{trial}
\end{equation}
We assume that the beam waist $w(z)$ is a known function and we 
consider only the position $r=0$,
so that the problem becomes one-dimensional in the axial coordinate $z$. 
Several simple approaches can be used to estimate $w(z)$ \cite{Couairon2007}.
Consistently with this approximation, we neglect the non-paraxial correction
$-i\omega_c^{-1}\partial_{\tau}\nabla _{\perp }^{2}$
(space-time focusing) in Eq. (\ref{eqgen}). One notes, however, that in
the special case where $w(z)$ is constant, the latter correction can
be get rid of simply by redefining the retarded time as 
$\tau \leftarrow \tau -\frac{2z}{k_0\omega_cw^2}$. 

With these simplifications, substituting Eq. (\ref{trial}) in Eq. (\ref{eqgen}) gives:
\begin{align}
& i\partial _{z}\frak{A} - \left( \frac{2}{k_0w^2}-
i\partial_z \ln(w) \right) \frak{A}
+\lambda |\frak{A}|^{2}\frak{A}-\eta
\frak{I}_K\frak{A} 
+i\gamma \partial _{\tau}\left( |\frak{A}|^{2}\frak{A}\right)
+i\delta \partial _{\tau }\left(\frak{I}_K\frak{A} \right)  \nonumber \\
& +i\frac{\beta }{2}|\frak{A}|^{2K-2}\frak{A}
+ i\epsilon \frak{I}_K\frak{A}=0
\label{eqpart}
\end{align}
where
\begin{equation}
\frak{I}_K(z,\tau)=\int_{-\infty }^{\tau }|\frak{A}(z,\tau^{\prime})|^{2K}d\tau ^{\prime } 
\end{equation}
and
\begin{equation}
\lambda=n_{2}k_{0},\;
\eta=\frac{k_{0}\alpha _{K}\rho _{m}}{2n^2_0\rho_{c}},\;
\gamma=\frac{2n_{2}}{c},\;
\delta=\frac{\alpha _{K}\rho_m}{2cn^2_0\rho_c},\;
\epsilon=\frac{1}{2}\sigma \alpha_K \rho_m
\end{equation}
We also use the estimate $\beta=\alpha _{K}\rho _{m}U_{I}$, 
where $U_{I}$ is the ionization potential.

\section{Energy balance}
From Eq. (\ref{eqpart}), it is straightforward to show that
\begin{equation}
\partial _{z} \left( w^2(z) \frak{I}_1(z,\infty) \right) =
- w^2(z)\left( \delta \frak{I}_{K+1}(z,\infty) + \beta \frak{I}_{K}(z,\infty) 
+ \epsilon \int_{-\infty}^{\infty} \rho |\frak{A}|^2d\tau \right)
\label{cons}
\end{equation}
where $w^2(z)\frak{I}_1(z,\infty)$ is proportional to the pulse energy at the position $z$. 
Equation (\ref{cons}) indicates that the $\delta$ term 
of Eq. (\ref{eqpart}) induces an energy leak. The relative
importance of the first two terms on the right hand side can be estimated by using the 
Gaussian form $|A|=A_{0}\exp
(-\tau ^{2}/T^{2})$, where $T$ is the pulse duration. Using the above
definitions, one obtains:
\begin{equation}
\frac{\delta \frak{I}_{K+1}(z,\infty) }{\beta
\frak{I}_{K}(z,\infty) }=\frac{\bar{\Phi}_{p}}{U_{I}}
\left(1+\frac{1}{K} \right)^{-1/2}
\label{ratio}
\end{equation}
where 
$\bar{\Phi}_{p}$ is the ponderomotive potential averaged over a laser cycle.
The $\delta$ term is thus related to the oscillation energy of the free electrons
in the laser field and 
this explains the energy leak induced by this term. 
For $K>1$, the energy leak due to
the $\delta $ term becomes comparable to that of the $\beta$ term 
when $\bar{\Phi}_{p}\approx U_{I}$. 
For the typical value $U_{I}\approx 10$ eV, this happens for the common
parameters $A_0^{2}\approx 10^{18}$ Wm$^{-2}$ and $\lambda_c \approx 1$ $\mu$m,
where $\lambda_c$ is the central wavelength.

It is worth noting that an alternative model to Eq. (\ref{eqgen}), involving the substitution
\begin{equation}
\left(1+ \frac{i}{\omega_c}\partial_{\tau} \right) \left(1-\frac{\nu_e}{\omega_c}\right)\rho A
\leftarrow \frak{T}^{-1}(\rho A)
\label{subst}
\end{equation} 
has also been used \cite{Gaeta2000,Zia2018}. 
With the latter substitution, Eq. (\ref{cons}), without the $\epsilon$ term 
now somehow included in the left hand side of (\ref{subst}), becomes
\begin{equation}
\begin{split}
\partial _{z} \left( w^2(z)\frak{I}_1(z,\infty) \right) & = - w^2(z) \left( -\delta \frak{I}_{K+1}(z,\infty) 
+2\delta\nu_e \int_{-\infty }^{\infty }|\frak{A}|^{2} \frak{I}_K d\tau  \right. \\
&\left. +\frac{2\delta \nu_e }{\omega_c}\int_{-\infty }^{\infty }|\frak{A}|^{2K+2}\partial_{\tau} \theta d\tau
+\beta \frak{I}_{K}(z,\infty) \right)
\end{split}
\label{cons2}
\end{equation}
The first term on the right hand side has a sign opposite to that in Eq. (\ref{cons}), 
implying that the $\delta $ term is now a source of energy. In that model, 
this source term can, however, 
be compensated by the negative second term for an appropriate choice of $\nu_e$. 
The third term has an undetermined sign
due to the $\partial_{\tau}\theta$, as discussed below.
\section{Perturbative solution}

To obtain a perturbative solution of the wave equation (\ref{eqpart}) we set
\begin{equation}
\frak{A}(z,\tau)=\psi(z,\tau)e^{i\theta(z,\tau)}
\label{ansatz}
\end{equation}
where $\psi$ and $\theta$ are real functions. Substituting Eq. (\ref{ansatz}) in Eq. (\ref{eqpart})
one obtains
\begin{eqnarray}
\partial _{z }\theta  + \frac{2}{k_0w^2}
-\lambda \psi ^{2}+\eta \frak{I}_K 
+\gamma \psi ^{2}\partial _{\tau }\theta +\delta \frak{I}_K \partial _{\tau }\theta &=&0
\label{realp} \\
\partial _{z }\psi +\partial_z \ln (w) \psi 
+3\gamma \psi ^{2}\partial _{\tau }\psi +\delta \left( \psi^{2K+1}+
\frak{I}_K \partial _{\tau }\psi \right) +\frac{\beta }{2}\psi ^{2K-1} +
\epsilon \frak{I}_k \psi &=&0
\label{imag}
\end{eqnarray}
for the real and imaginary parts, respectively. Neglecting dispersion allowed decoupling 
$\psi$ from $\theta$ in Eq. (\ref{imag}).
Note that dispersion was taken into account by Tzoar and Jain in their
perturbative treatment of the propagation of laser pulses in optical fibres,
but when retaining only the additionnal $\gamma$ term \cite{Tzoar1981}. 

We seek a perturbative solution for $\psi$ by substituting the first
order series expansion
\begin{equation}
\psi =\psi _{0}+\gamma \psi _{\gamma}+\delta \psi _{\delta}+\beta \psi _{\beta}+ \epsilon \psi_{\epsilon}
\label{seriespsi}
\end{equation}
in Eq. (\ref{imag}). 
The resulting equation holds for arbitrary values of $\gamma$, $\delta$, $\beta$ and $\epsilon$ provided
\begin{eqnarray}
\psi _{0}(z,\tau)&=&\frac{w_0}{w(z)}F(\tau)     
\label{psi0} \\
\psi _{\gamma}(z,\tau) &=& -\frac{w_0}{w(z)}\phi_1(z_0,z) \partial _{\tau }F^{3}(\tau)   \\
\psi _{\delta}(z,\tau) &=&-\frac{w_0}{w(z)} \phi_K(z_0,z) \left( F^{2K+1}(\tau)+ 
\partial _{\tau }F(\tau) \xi_K(\tau) \right) \\
\psi_{\beta}(z,\tau) &=& -\frac{1}{2} \frac{w_0}{w(z)}\phi_{K-1}(z_0,z)F^{2K-1}(\tau) \\
\psi_{\epsilon}(z,\tau) &=& -\frac{w_0}{w(z)} \phi_K(z_0,z)F(\tau)\xi_K(\tau)
\end{eqnarray}

In these expressions, $w_0=\min(w(z))$ within the range of $z$ considered, and
\begin{equation}
\phi_N(z_0,z)=\int_{z_0}^{z} \left(\frac{w_0}{w(z^{\prime})}\right)^{2N}dz^{\prime},\;\;\;
\xi_N(\tau)=\int_{-\infty}^{\tau}F^{2N}(\tau^{\prime})d\tau^{\prime}
\label{integrale}
\end{equation}
where $z_0$ is the coordinate $z$ where the initial condition is defined.

Assuming that the zero order temporal pulse shape is the Gaussian function
\begin{equation}
F(\tau )=A_0e^{-\tau ^{2}/T^{2}}
\label{gauss}
\end{equation}
one finds the simple closed-form expression for the pulse amplitude
\begin{equation}
\begin{split}
\psi (z ,\tau ) =A_{0}\frac{w_0}{w(z)}e^{-\tau ^{2}/T^{2}}
& \left\{ 1+\frac{6\gamma A_{0}^{2} }{T} \phi_1(z_0,z) \frac{\tau }{T}e^{-2\tau ^{2}/T^{2}}   \right. \\
& \left. -\delta A_{0}^{2K}\phi_K(z_0,z) \left[ e^{-2K\tau ^{2}/T^{2}}-\sqrt{\frac{\pi }{2K}}
\frac{\tau }{T}\left( 1+\text{erf} \left( \sqrt{2K}\frac{\tau }{T}\right)
\right) \right] \right. \\
& \left. -\frac{\beta }{2} A_{0}^{2K-2}\phi_{K-1}(z_0,z)e^{-\left( 2K-2\right) \tau^{2}/T^{2}} \right. \\
& \left. -\epsilon A_0^{2K} \phi_K(z_0,z)\sqrt{\frac{\pi }{8K}}T 
\left( 1+\text{erf} \left( \sqrt{2K}\frac{\tau }{T} \right) \right)
\right\}
\end{split}
\label{pertpsi}
\end{equation}

In the particular case where $w(z)=w_0$, and only the $\gamma$ term is retained, 
Anderson and Lisak found that the exact solution of Eq. (\ref{imag})
for the zero order temporal profile, Eq. (\ref{gauss}), is the solution 
of the algebraic equation \cite{Anderson1983}:
\begin{equation}
\psi=A_{0}\exp \left( -\left( \tau -3\gamma (z-z_0) \psi ^{2}\right)
^{2}/T^{2}\right)
\label{exact}
\end{equation}
The solution of Eq. (\ref{exact})
has a maximum of $\psi=A_0$ at $\tau= \mu T$, where 
$\mu =3\gamma (z-z_0) A_{0}^{2}/T$
is the scaling parameter, but tends to $\psi_0$ for small values of $\psi$. 
It thus becomes very steep past the maximum when $\mu$ is on the order of 1 or 
greater \cite{Anderson1983}.
The series expansion of the solution of Eq. (\ref{exact}) up to the order $\mu ^{2}$ is:
\begin{equation}
\psi = A_0 e^{-\tau ^{2}/T^{2}}\left[ 1+2\mu \frac{\tau }{T}
e^{-2\tau ^{2}/T^{2}}-\mu ^{2}\left( 1-10\frac{\tau ^{2}}{T^{2}}\right)e^{-4\tau ^{2}/T^{2}} 
+O\left(\mu^3\right) \right] 
\label{series}
\end{equation}

Consistently with the exact solution, the first order term in 
$\mu$ in Eq. (\ref{series}), which coincides with that
of Eq. (\ref{pertpsi}), shifts the peak to the back of the pulse ($\tau >0$).
The $\mu^2$ correction lowers the amplitude around $\tau=0$ and shifts the peak 
somewhat further to the back.

From Eq. (\ref{pertpsi}), one can identify the following dimensionless scaling parameters
for the four perturbation terms
\begin{equation}
3\gamma A_{0}^{2}\phi_1(z_0,z) /T,\; \delta A_{0}^{2K}\phi_K(z_0,z),\;
\beta A_{0}^{2K-2}\phi_{K-1}(z_0,z) /2,\; \epsilon A_0^{2K}T\phi_K(z_0,z)\sqrt{\pi/2K}
\label{scaling}
\end{equation}

\begin{figure}
\includegraphics[scale=0.4]{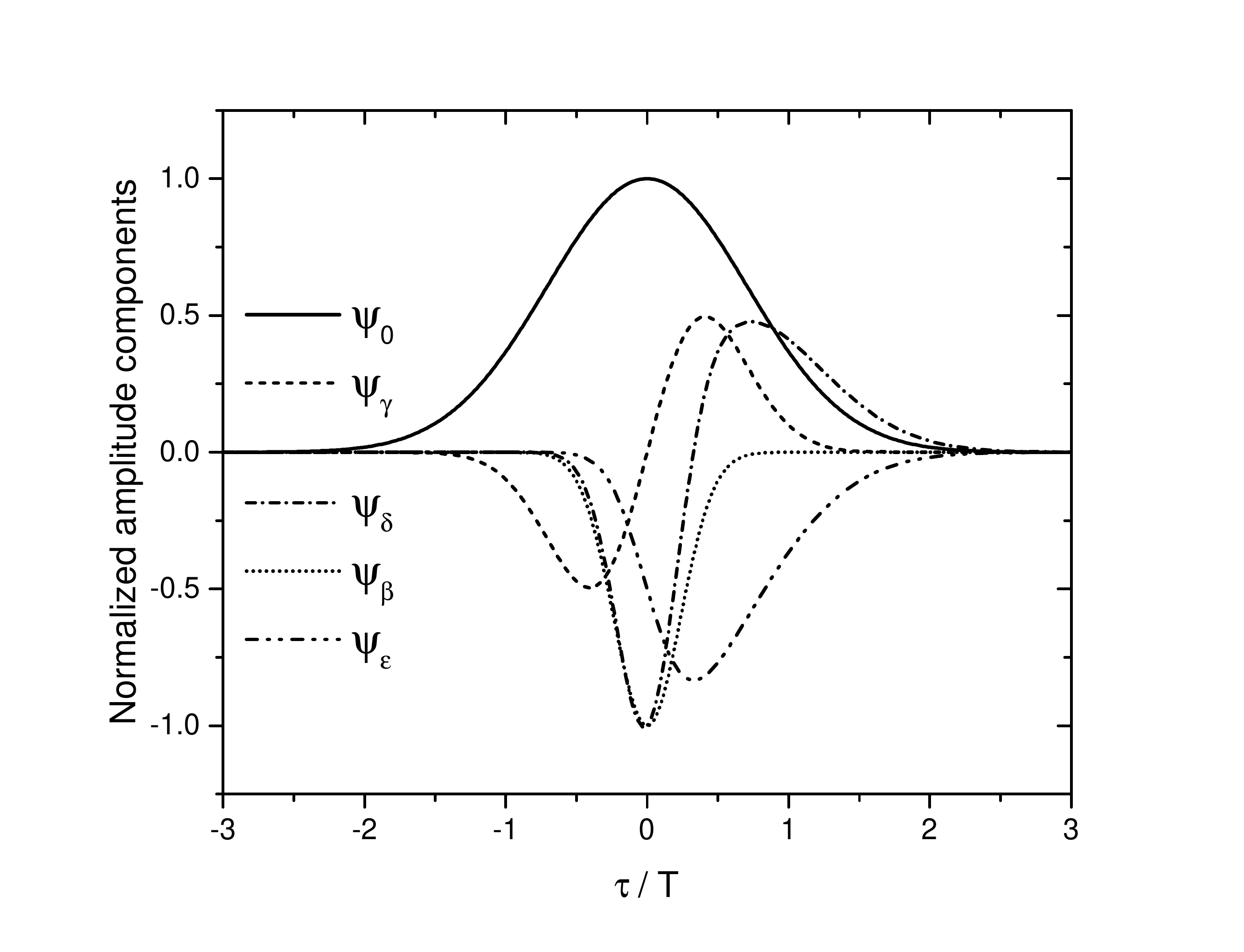} 
\caption{Time dependence of the terms contributing to the perturbative 
solution Eq. (\ref{pertpsi}) for $K=5$.}.
\label{Fig1}
\end{figure}

The time dependence of the five terms of Eq. (\ref{pertpsi}) 
are illustrated in Fig. \ref{Fig1} for $K=5$ 
with the scaling parameters (\ref{scaling}) set equal to 1. As expected,
the $\beta$ term (absorption due to optical field ionization) induces 
a symmetric depletion at 
the initial peak of the pulse where the ionization rate is maximum. 
The $\gamma $ term (Kerr effect) discussed above
produces an antisymmetric contribution consistently with its general energy
conserving property (although this property holds only to the first order in 
$\gamma$ in the current level of approximation). The $\epsilon$ term (collisional absorption) 
produces an off-centred asymmetric depletion
as this effect is most efficient near the peak of the pulse and where the electron density is highest.
More interestingly, the $\delta $ term (time derivative of optical field ionization) brings an
asymmetric contribution, with a sharp depletion at $\tau =0$, and a shift of
the peak toward the back of the pulse. With reference to the discussion of Section III, 
this asymmetric shape can be interpreted as an absorption
of the pulse energy by the free electrons at the front and the center of the pulse, and
a partial restitution of the electron energy at the back of the pulse as the 
amplitude decreases. In
general, one thus expects a shift of the peak toward the back of the
pulse and a steepening effect due to both the $\gamma$ and the $\delta$
terms. This effect is however hindered by the $\epsilon$ term, which tends to deplete
the amplitude at the back of the pulse.

The phase $\theta (z ,\tau )$ is obtained by substituting the first order series expansions
\begin{equation}
\theta =\theta _{0}+\gamma \theta _{\gamma}+\delta \theta _{\delta}
+\beta \theta_{\beta} +\epsilon \theta_{\epsilon}
\label{seriestheta}
\end{equation}
and (\ref{seriespsi}) in Eq. (\ref{realp}). One obtains
\begin{align}
\theta _0(z,\tau) &= -\frac{2}{k_0w_0^2}\phi_1(z_0,z) + \lambda \phi_1(z_0,z) F^{2}(\tau)
-\eta \phi_K(z_0,z) \xi_K(\tau) \\
\theta _{\gamma}(z,\tau) & =-2\lambda \Phi_{1,1}(z_0,z) \partial_{\tau }F^{4}(\tau)
+ \eta \left( \frac{6K}{2K+2} \Phi_{K,1}(z_0,z) + \Phi_{1,K}(z_0,z) \right) F^{2K+2}(\tau) \\
\theta _{\delta}(z,\tau) & =-2\lambda \Phi_{1,K}(z_0,z) F^{2K+2}(\tau) 
-\lambda ( \Phi_{1,K}(z_0,z)+\Phi_{K,1}(z_0,z) ) \partial_{\tau} F^2(\tau) \xi_K(\tau) \nonumber \\
& +\eta (2K-1) \Phi_{K,K}(z_0,z) \xi_{2K}(\tau) 
+ 2\eta \Phi_{K,K}(z_0,z) F^{2K}(\tau)\xi_K(\tau) \\
\theta _{\beta}(z,\tau) & = -\lambda \Phi_{1,K-1}(z_0,z) F^{2K}(\tau) 
+\eta K \Phi_{K,K-1}(z_0,z) \xi_{2K-1}(\tau) \\
\theta_{\epsilon}(z,\tau) & =-2\lambda \Phi_{1,K}(z_0,z) F^2(\tau) \xi_K(\tau) 
+ 2\eta K \Phi_{K,K}(z_0,z)\int_{-\infty}^{\tau} F^{2K}(\tau^{\prime})\xi_K(\tau^{\prime})d\tau^{\prime} 
\end{align}
where

\begin{equation}
\Phi_{i,j}(z_0,z)= \int_{z_0}^z \left( \frac{w_0}{w(z^{\prime})} \right)^{2i} \phi_j(z^{\prime}) dz^{\prime}
\end {equation}

The full expression of $\theta(z,\tau)$ using Eq. (\ref{gauss}) is quite 
lengthy and will not be developed here for that reason.
One notes that the first term of $\theta_0(z,\tau)$ is the Gouy phase for the 
selected waist function $w(z)$. 

More interesting than the phase itself is the instantaneous frequency shift $\Delta \omega(z,\tau)
=\omega(z,\tau) -\omega_c=-\partial_{\tau} \theta(z,\tau)$ where $\omega(z,\tau)$ is the frequency
generated by the nonlinear effects \cite{Boyd2003}. 
For the time profile (\ref{gauss}), one finds for the component
corresponding to $\theta_0$,
\begin{equation}
\Delta \omega _{0} =\frac{4\lambda A_{0}^{2}}{T}\phi_1(z_0,z)\frac{\tau }{T}e^{-2\tau
^{2}/T^{2}}+\eta A_{0}^{2K}\phi_K(z_0,z)e^{-2K\tau^{2}/T^{2}}
\label{om0}
\end{equation}

In this expression, the first term produces a red shift at the front of the pulse due to the increase 
of the Kerr component of the index of refraction $n_{Kerr}=n_0+n_{2}|A|^{2}$ 
and a blue shift at the back due
to the subsequent decrease of $n_{Kerr}$, resulting in the well-known linear chirp around
$\tau =0$. The second term brings a narrower blue shift contribution around the center of the pulse
due to the generated electrons, which decrease the index of refraction as 
$n_{elec} \approx 1 - \rho /2 \rho_c$. 
The effect of the remaining components of $\Delta \omega$ will be illustrated in the following example.

\section{Example: laser pulse focused in air}
\label{Ex}

As an example, we consider a short laser pulse, with a duration $T=50$ fs 
(or a full width at half maximum of the intensity of $T\sqrt{2\ln{2}}=59$ fs)
and a central wavelength $\lambda_c = 800$ nm, focused in
ambient air, where the molecule density is $\rho _{m}=2.7\times 10^{25}$ m$^{-3}$,
with $20\%$ O$_2$, and $n_0=1$. 
The waist will be assumed of the form
$w(z)=w_0\left(1+\left(z/z_R^{eff} \right)^2\right)^{1/2}$, where $w_{0}=50$ $\mu $m, 
and $z_R^{eff}$ is the effective Rayleigh length.
We consider $A_0^2 \leq 1.4 \times 10^{18}$ Wm$^{-2}$  so that
the peak intensity $\psi^2$ at focus $z=0$ is lower than that.

For these parameters, the energy of a Gaussian pulse is $E_0 \lesssim 0.34$ mJ, and its power
$P_0 \approx A_{0}^{2} \frac{\pi}{2} w_{0}^{2}$
$\lesssim  5.5$ GW is below the critical power for self focusing
$P_{cr}\approx \lambda ^{2}/2\pi n_{2}\approx 12.7$ GW, where we used 
$n_{2}=0.8 \times 10^{-23}$ m$^{2}$ W$^{-1}$ \cite{Zahedpour2015}. 
We estimate the importance of dispersion from the relative time broadening 
$\Delta T/T \sim 2\left( k^{\prime \prime} z /T^{2}\right)^{2}$ of the pulse. 
Using $k^{\prime \prime} = 2 \times10^{-29}$ s$^{2}$ m$^{-1}$ for ambient 
air \cite{Couairon2007}, one finds that, for the
propagation distance $z =1$ m, $\Delta T/T$ is on the order of $10^{-4}$,
which is negligible as assumed above.
Ionization of O$_{2}$ is more effective than that of N$_{2}$ since their ionization
potentials $U_{I}$ are 12.1 eV and 15.6 eV, respectively. The Keldysh
parameter $\gamma _{K}=\omega A_{0}\sqrt{U_{I}/c\varepsilon _{0}}\approx
10^{-4}$ indicates that ionization is in the tunnel regime. The coefficients 
$K$ and $\alpha _{K}$ of the ionization rate, fitted as $W=\alpha
_{K}|\psi|^{2K}$ for O$_{2}$, can be estimated from the PPT model as $K\approx 5$ 
and $\alpha _{K}\approx 5 \times 10^{-79}$ m$^{10}$ W$^{-5}$ s$^{-1}$ around 
$\psi^2=10^{18}$ Wm$^{-2}$ \cite{Schwarz2012} producing an ionization rate
$W \approx 5 \times 10^{11}$ s$^{-1}$ and an electron density near the peak of the pulse
$\rho_{peak} \approx 4 \times10^{22}$ m$^{-3}$, which is much smaller than the
critical plasma density $\rho _c=1. 7 \times 10^{27}$ m$^{-3}$. Considering few eV
electrons and an electron-neutral geometrical cross section of $\sim 10^{-19}$ m$^2$, 
one obtains the ratio $\nu_e/\omega_c \sim 10^{-3}$. Therefore, 
$\sigma \approx 4.5 \times 10^{-24}$ m$^2$.

We estimate the effective Rayleigh length as 
$z_R^{eff} \approx \left(w_0/\partial^2_{z}w|_{w=w_0} \right)^{1/2}$, where \cite{Schwarz2012}
\begin{equation}
\partial^2_{z}w|_{w=w_0} \approx \frac{4}{k^2w_0^3}\left(1-\frac{P_0}{P_{cr}}\right)
+ \frac{2K}{(K+1)^2w_0}\frac{\rho_{peak}}{\rho_{cr}}
\label{second}
\end{equation}

Using the above parameters, one finds for $\psi^2 = 10^{18}$ Wm$^{-2}$, 
$z_R^{eff} \approx 1$ cm, which is nearly the same value as the Rayleigh lenght in vacuum 
$z_R = kw_0^2/2=0.98$ cm,
which is the value used in the following.
This result can be understood as a near cancellation of Kerr self-focusing
and plasma defocusing since their scalelengths are
$1/k_0n_2A_0^2 \approx 1.3$ cm and 
$2\rho_c/k_0\rho_{peak} \approx 1.1$ cm, respectively \cite{Couairon2007}.

We assume that, well before the focus ($z_0 \ll -z_R$), 
$\frak{A}(z_0,\tau)=\psi_0(z_0,\tau)$, where $\psi_0$ 
is given by Eqs. (\ref{psi0}) and (\ref{gauss}), 
and we are interested in the field envelope $\frak{A}(z,\tau)$ 
well past the focus ($z \gg z_R$). All the nonlinear effects take place around the focus
where the intensity $\psi^2$ is maximum.

For the selected function $w(z)$, the integrals $\phi_i$ and $\Phi_{i,j}$ 
evaluated between $-\infty$ and $\infty$ can be calculated exactly. 
They are on the order of $z_R$ and $z_R^2$, respectively.
Using the above parameters, one finds that the dimensionless scaling parameters identified in
Eq. (\ref{scaling}) are, for$A_0^2=1.2 \times 10^{18}$ Wm$^{-2}$:
$3\gamma A_{0}^{2}\phi_1 /T = 0.118 $, 
$\delta A_{0}^{10} \phi_5 = 0.054$, $\beta A_0^8 \phi_4/2 = 0.052$,
and $\epsilon A_0^{10} \phi_5 \sqrt{\pi/10} = 0.018$. 
Since all these parameters (and in particular the sum of the $\delta$ and $\beta$ parameters)
are much smaller than 1, the laser pulse can be considered
in the perturbative regime. 
For $A_0^2=1.4 \times 10^{18}$ Wm$^{-2}$:
$3\gamma A_{0}^{2}\phi_1 /T = 0.138 $, 
$\delta A_{0}^{10} \phi_5 = 0.117$, $\beta A_0^8 \phi_4/2 = 0.097$,
and $\epsilon A_0^{10} \phi_5 \sqrt{\pi/10} = 0.038$. In that case the scaling
parameters are larger and might depart from the perturbative regime.
\begin{figure}
\includegraphics[scale=0.4]{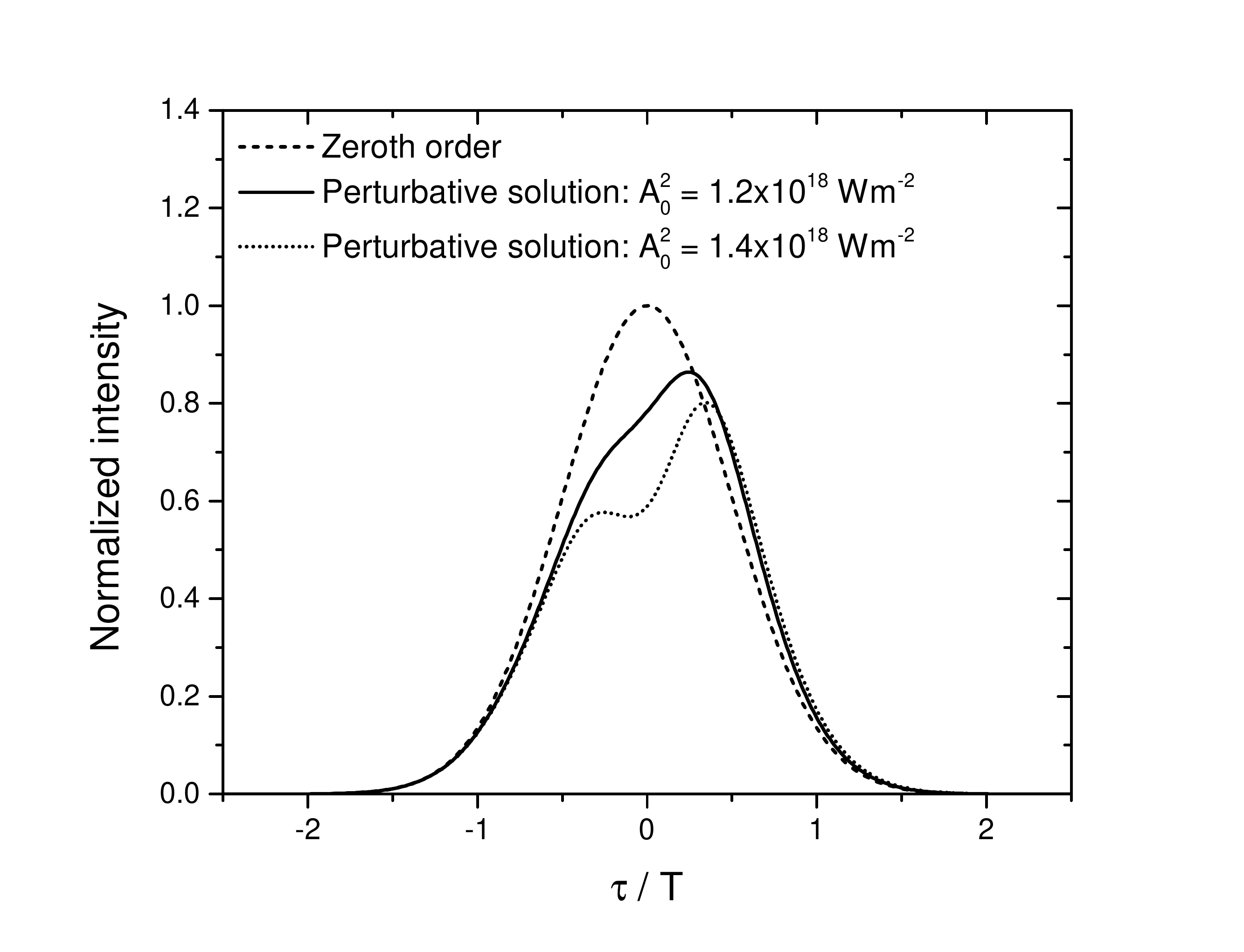} 
\caption{Normalized pulse intensity of the zeroth order ($z \ll -z_R$) and of the
full perturbative solution at $z \gg z_R$.}.
\label{Fig2}
\end{figure}

The shape of the pulse intensity $\psi^2$ normalized as 
$\left(\frac{w(z)}{w_0}\frac{\psi}{A_0}\right)^2$ 
is illustrated in Fig. \ref{Fig2} for the zeroth order $\psi=\psi_0$ ($z \ll -z_R$)
and for the full perturbative solution at $z \gg z_R$ for
$A_0^2=1.2 \times 10^{18}$ Wm$^{-2}$ and $1.4 \times 10^{18}$ Wm$^{-2}$. 
It is worth stressing
that the results depend on $w(z)$ only through the integrals $\phi_i$ and $\Phi_{i,j}$ 
evaluated between $-\infty$ and $\infty$.
According to Fig. \ref{Fig1} and to the scaling parameters evaluated above, 
the $\gamma$, $\delta$, $\beta$, and $\epsilon$ terms have comparable 
magnitudes but act at different times.
Consistently with Fig. \ref{Fig1}, the perturbative solution, as compared to the
zeroth order, is depleted at the front and at the centre of the pulse due to
electron generation and their energy absorption, and enhanced at the back mostly due to the 
Kerr effect ($\gamma$ term).
As $A_{0}^{2}$ is increased, the shoulder around $\tau=0$ becomes more and more pronounced and,
for $A_0^2 \gtrsim 1.4 \times 10^{18}$ Wm$^{-2}$, forms a hole 
in the intensity profile. The intensity profiles obtained share much resemblance with those of the numerically 
calculated \textit{light bullets} propagating in dielectric materials \cite{Zia2018,Smetanina2013}, 
where the Kerr effect is however much larger than in air. In the present case,
the normalized pulse shape stops changing due to the divergence of the laser pulse 
past the focus.
 
\begin{figure}
\includegraphics[scale=0.4]{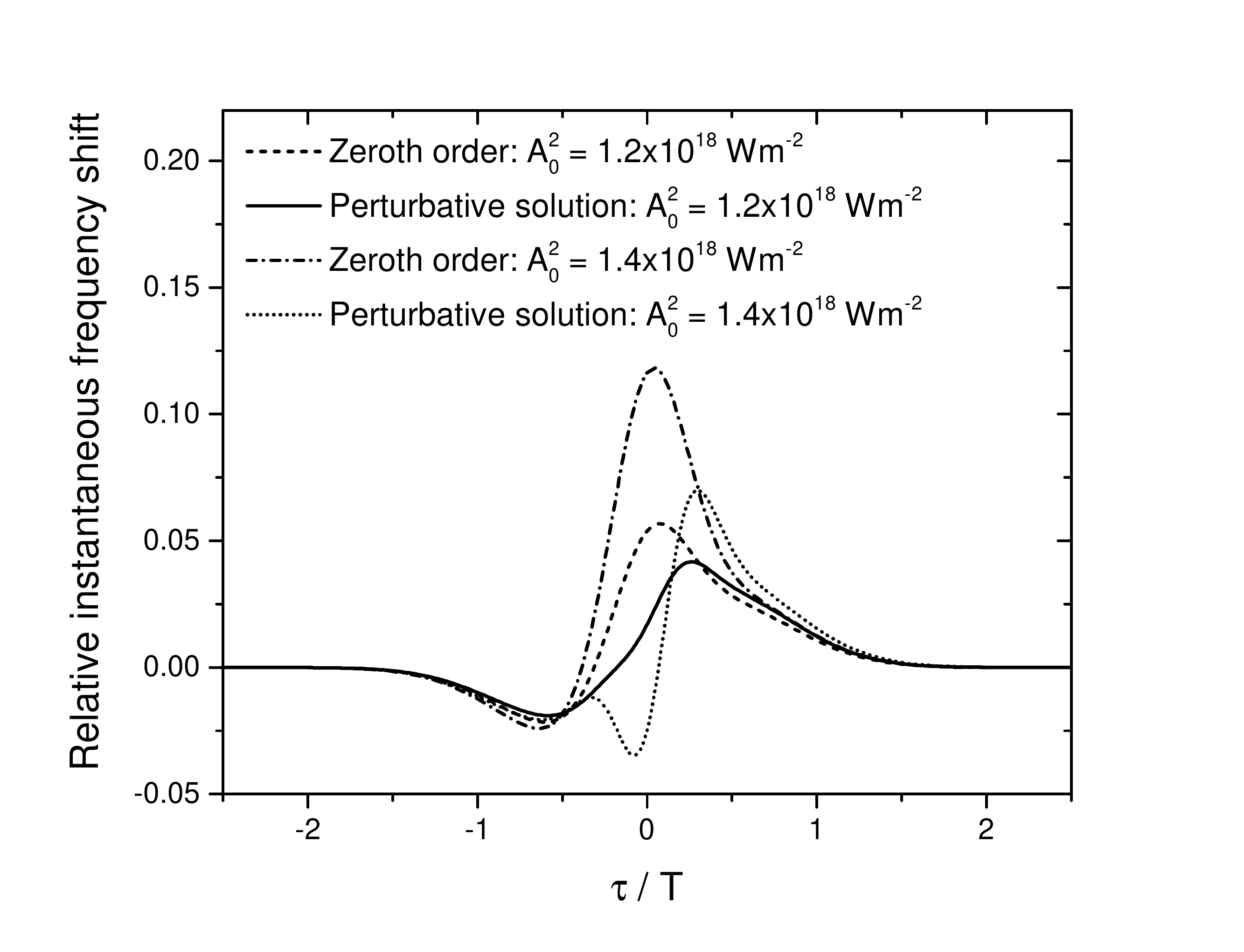} 
\caption{Relative instantaneous frequency shift  $\Delta \omega/\omega_c$ 
and the zeroth order contribution at $z \gg z_R$.}
\label{Fig3}
\end{figure}

The relative frequency shift $\Delta \omega /\omega_c$ 
is shown in Fig. \ref{Fig3} as well as the zeroth order solution 
$\Delta \omega_0 /\omega_c$, given by Eq. (\ref{om0}), for $z \gg z_R$.
For $A_{0}^{2} = 1.2 \times 10^{18}$ Wm$^{-2}$ 
the full frequency shift is clearly dominated by the zeroth order contribution
for all values of $\tau$. 
The perturbative terms $\delta$, $\beta$, and $\epsilon$
flatten the frequency shift around $\tau=0$.
When $A_{0}^{2} \gtrsim 1.4 \times 10^{18}$ Wm$^{-2}$, 
$\Delta \omega /\omega_c$
shows complicated oscillations due to the sharp negative 
contributions of the perturbative terms near $\tau=0$.
The total frequency shift is nevertheless always dominated by the 
$\lambda$ term (Kerr effect) in
$\Delta \omega _0$ at the front  (red shift) and at the 
back (blue shift) of the pulse due to the large broadeness of $F^2(\tau)$. 
Note that, for $A_{0}^{2} = 1.2 \times 10^{18}$ Wm$^{-2}$, 
the dimensionless scaling parameters of the
zeroth order phase component $\theta_0$ are
$\lambda A_0^2\ \phi_1=2.33$ and $\eta A_0^{10} \phi_{5}T\sqrt{\pi/10}=3.56$,
and thus cannot be considered as perturbative components.

\begin{figure}
\includegraphics[scale=0.4]{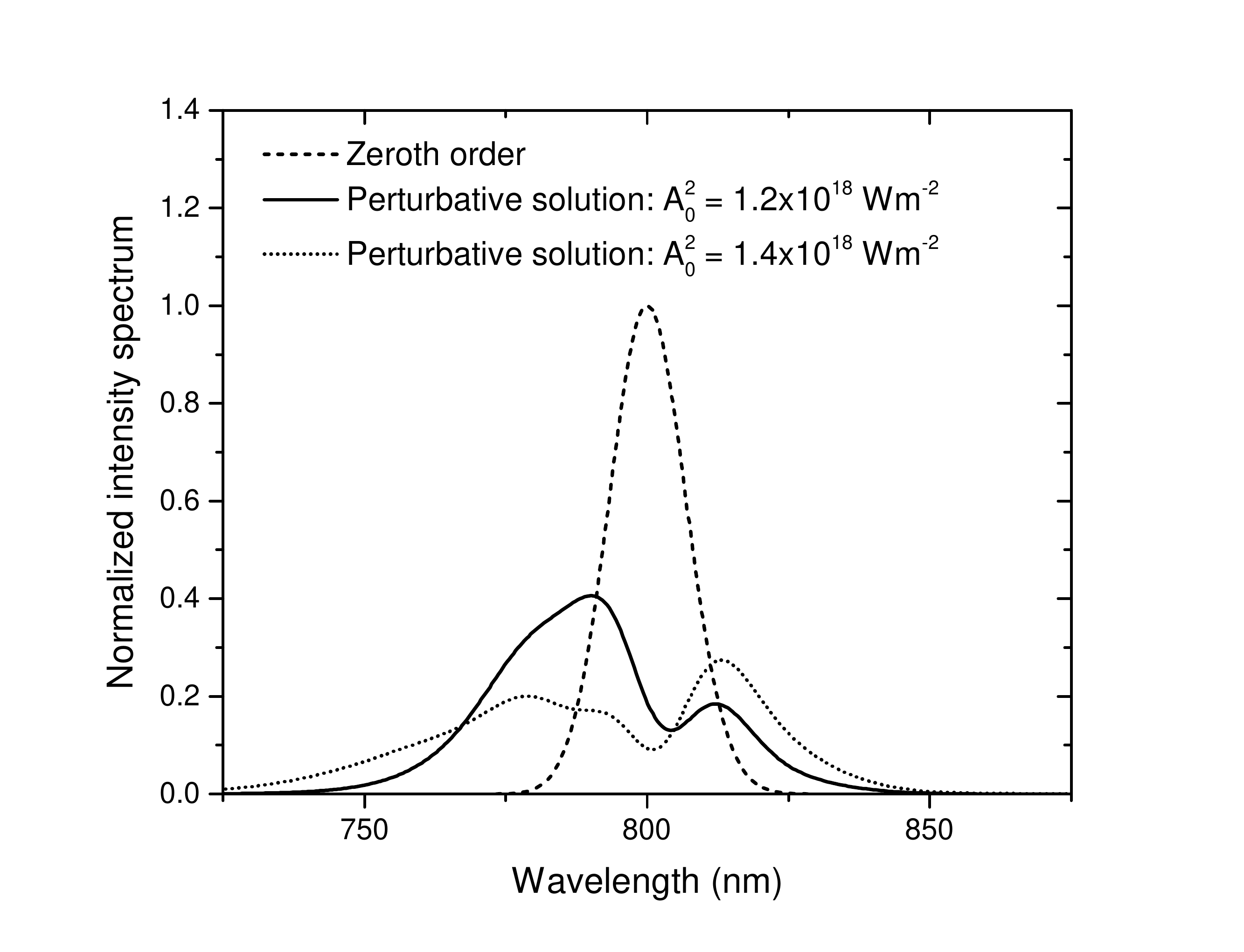} 
\caption{Intensity spectrum around the central wavelength $\lambda_c =800$ nm at $z \gg z_R$
and the zeroth order spectrum ($z \ll z_R$).}
\label{Fig4}
\end{figure}

Finally, we calculate the intensity spectrum of the pulse, here normalized as
\begin{equation}
\tilde{I}(z,\omega) = \frac{1}{\pi A_0^2} \left(\frac{w(z)}{w_0} \right)^2 
\left | \int_{-\infty }^{\infty}\frak{A}(z,\tau) e^{i\omega \tau} d\tau \right |^2 
\label{fourier}
\end{equation}
since it is a directly measurable quantity. Figure \ref{Fig4} shows the 
intensity spectra as a function
of wavelength for $z \gg z_R$ as well as the zeroth order spectrum ($z \ll -z_R$). 
The full perturbative spectra are strongly broadened and depleted as compared to 
the zeroth order spectrum. Consistently with the discussion around Fig. \ref{Fig3},
the spectrum for  $A_0^2=1.2 \times 10^{18}$ Wm$^{-2}$ undergoes mostly 
a blue shift and a lower peak appears on the red side.
The peak shift obtained is $-10$ nm from the initial value $\lambda_c=800$ nm. 
One notes that a similar double-humped spectrum for a tightly focused laser pulse in nitrogen was recently 
measured in a regime which is however not clearly perturbative \cite{Clerici2019}.
When $A_{0}^{2}$ is increased, the two peaks become more and more 
comparable in height and the shorter wavelength peak broadens and shifts further to the blue side.

\section{Summary}
We have employed a perturbative approach to derive closed-form expressions for the solution of the wave equation discussed by Berg\'{e} et al. \cite{Berge2007}, describing the envelope amplitude and phase of a laser pulse propagating in optically transparent media. The the waist $w(z)$ of the laser pulse was assumed to be known so that the problem becomes one-dimensional in the axial
coordinate $z$.
The perturbative approach employed in this work allowed to display the contribution of the various terms of the basic model equation to the temporal shape of the pulse in the limit where the nonlinear effects are modest.
The most interesting term is likely
$\partial _{\tau }\left( \rho A\right)$ in Eq. (\ref{eqpart}) (the $\delta $ term), associated with field ionization, which was shown to produce a depletion of the initial peak amplitude and its displacement toward the back of the pulse. This term was also shown to produce an overall energy transfer toward the free electrons oscillating in the laser field. 
Moreover, dimensionless parameters enabling to size the nonlinear effects have been extracted from the perturbative solution. The solution obtained was illustrated on the specific example of a short laser pulse focused in air using common laser parameters. 
Experiments prepared in appropriate perturbative conditions could be performed to verify some conclusions of this work.

\newpage
\bibliography{biblio}

\end{document}